\documentclass[twocolumn,showpacs,aps,prl]{revtex4}
\usepackage{graphicx,amsmath,amssymb}

\begin{document}
\title{Experimental demonstration of classical Hamiltonian monodromy in the 1:1:2 resonant elastic pendulum}
\author{N. J. Fitch$^1$, C. A. Weidner$^1$, L. P. Parazzoli$^1$, H. R. Dullin$^2$, H. J. Lewandowski$^1$}
\affiliation{$^1$JILA and Department of Physics,
University of Colorado, Boulder, Colorado 80309-0440 \\
$^2$School of Mathematics and Statistics,
The University of Sydney,
Sydney, NSW 2006, Australia}

\date{\today}
\begin{abstract}
The 1:1:2 resonant elastic pendulum is a simple classical system that displays the phenomenon known as Hamiltonian monodromy.  With suitable initial conditions, the system oscillates between nearly pure springing and nearly pure elliptical-swinging motions, with sequential major axes displaying a stepwise precession.  The physical consequence of monodromy is that this stepwise precession is given by a smooth but multivalued function of the constants of motion.  We experimentally explore this multivalued behavior.  To our knowledge, this is the first experimental demonstration of classical monodromy.
\end{abstract}

\maketitle

After more than three hundred years since the formulation of Newton's laws of motion, one would expect that a system as simple as a mass on a spring would have been fully understood for some time.  In fact, an in-depth investigation of even a subset of its possible dynamics produces a number of surprises.  Chief among these is a phenomena known as Hamiltonian monodromy, which was introduced by Duistermaat in 1980 as a topological obstruction to the existence of global action-angle variables \cite{Duistermaat80}.  In the resonant elastic pendulum, monodromy has easily observable physical consequences.  Specifically, the observed stepwise precession of the elliptical swinging major axis is given by a smooth, but {\it multivalued} function of the constants of motion.  This functional form results in loops of values of the constants of motion having differing overall behavior, depending on the loop's topology.

If monodromy were limited to the resonant elastic-pendulum system as a special case, it would be considered just an esoteric detail.  However, it has been shown theoretically to exist in many other common and relatively simple systems, including the spherical pendulum, the Lagrange top, and the Kirchoff top \cite{Duistermaat80,Bates93,vivolo03,Cushman97}.  Most intriguing are the quantum mechanical implications in atomic and molecular systems. When a classical system exhibits monodromy, the energy eigenstates of the corresponding quantum system can not be mapped onto a simple lattice labeled by integer quantum numbers. Defects in the lattice of eigenstates are the striking signature of monodromy in the global structure of a quantum spectrum. In addition to the static effect that monodromy has on global quantum numbers, dynamical consequences have also recently been predicted \cite{Delos08a,Delos08b}.  Important quantum systems have been shown theoretically to have monodromy, including ellipsoidal billiards \cite{Waalkens02a}, trapped Bose gases \cite{Waalkens02b}, the $H_{2}^{+}$ molecular ion \cite{Waalkens04}, the hydrogen atom in combined electric and magnetic fields \cite{Cushman00,Schleif07,Efstathiou08},  dipolar symmetric-top molecules in electric fields \cite{Kozin03},  and the ro-vibrational spectra quasi-linear molecules such as $CO_{2}$ \cite{Giacobe04,Child99,Cushman04}.  As we discuss below, monodromy can occur near relative equilibria. There is thus the intriguing possibility that the singular behavior of monodromy may be a common feature of dynamics near chemical isomerization thresholds \cite{Zhang07}.

\begin{figure}
\includegraphics[width=3.5cm,height=3.0cm]{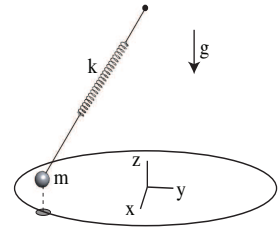}
\caption{Diagram of an elastic pendulum in a gravitational field of acceleration g.  One type of motion is swinging with fixed energy in the z-dimension.  This motion projected onto the XY plane is an ellipse.}
\label{system}
\end{figure}

A quantum analog of the resonant elastic pendulum under consideration here (Fig.\ref{system}) is the Fermi resonance in the $CO_{2}$ molecule, whose monodromatic features have been thoroughly investigated theoretically \cite{Cushman04,Giacobe04}. Despite the large number of systems in which monodromy is theoretically predicted, there have been no previous classical experiments and only a single quantum experiment \cite{Winnewisser05} of which we are aware. In developing a more heuristic understanding of monodromy in quantum systems, it is useful to have a classical example to guide one's intuition. Thus we designed our experiment on a readily realized classical system in which the consequences of monodromy are relatively easy to observe.

Hamiltonian monodromy is a property of certain integrable systems.  For a more complete introduction than appears here, see Refs. \cite{Efstathiou04,Sadovskii06}.
For concreteness, consider an integrable conservative classical system with N degrees of freedom described by the Hamiltonian $H(q_i,p_i)$ with generalized coordinates $\{q_{i}\}$ and conjugate momenta $\{p_{i}\}$, with $i=1,\ldots,N$.
The $\{q_{i},p_{i}\}$ form the coordinates of the 2N dimensional phase space $P^{2N}_{q,p}$.  Such systems contain a set of N independent constants of motion $\{F_{k}(q_{i},p_{i})\}$  that have mutually vanishing Poisson Brackets, $\{F_{m},F_{n}\}_{PB}=0$ for all $m,n$.  The independence of the $F_{k}$ refers to the linear independence of the differentials $dF_{k}$ almost everywhere in $P^{2N}_{q,p}$.  These types of systems allow the construction of local action-angle variables \cite{Arnold88}.  These variables are a set of coordinates (action-angles) and conjugate momenta (actions) that obey particularly simple equations of motion.  Specifically, the actions are all conserved quantities whose conjugate coordinates are $2 \pi$ periodic and evolve linearly in time with constant frequencies.  Ratios of these frequencies, known as {\it rotation numbers}, convey information about the local choice of action-angle variables.  If a rotation number can not be defined globally by a smooth function, then neither can the action-angle variables used to calculate it.

\begin{figure}
\includegraphics[width=7.5cm,height=3.5cm]{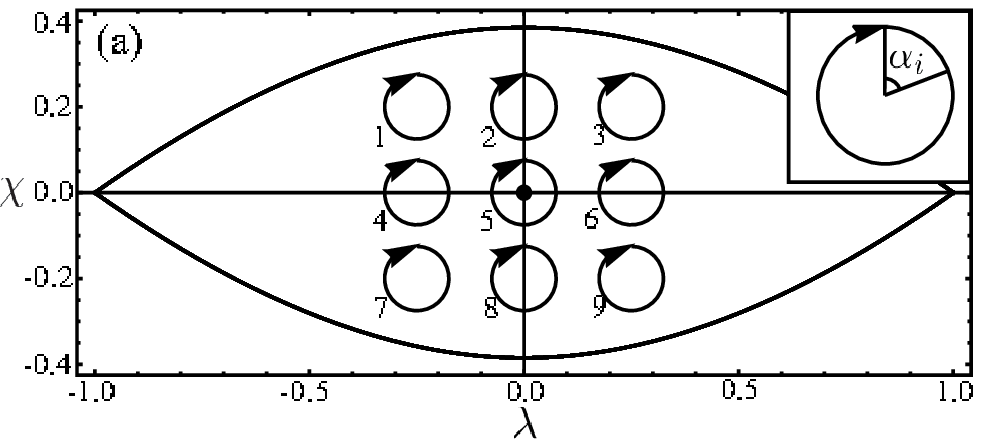}
\includegraphics[width=8cm,height=4.0cm]{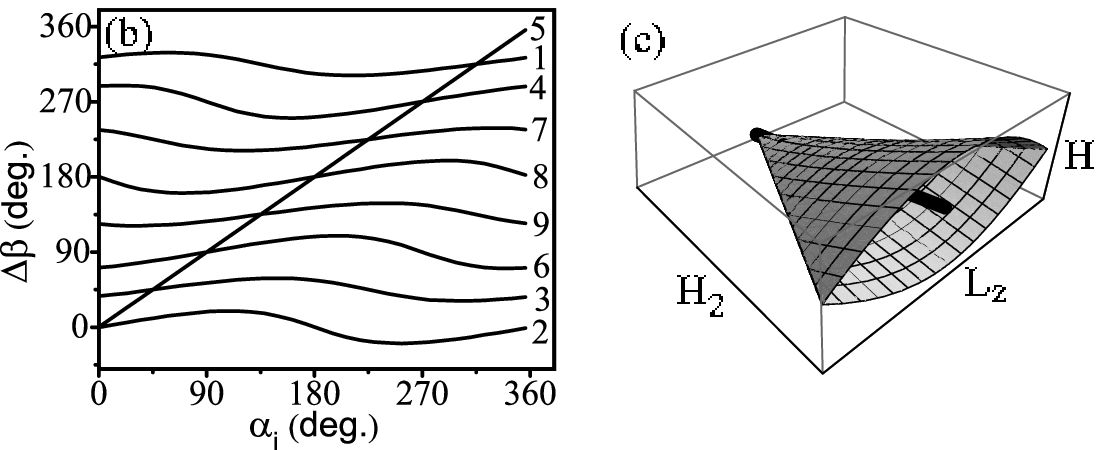}
\caption{(a) Loops of constants of motion in ($\lambda,\chi$) space generated from various sets of initial conditions.  The singularity at the origin corresponds to pure springing motion. (b) Resulting swing-plane precession angles for loops corresponding to those in (a), to lowest order in ($\lambda$, $\chi$).  Position on a loop is labeled by the angles $\alpha_{i}$, which are defined from the center of the $i^{th}$ loop with respect to the vertical.  Clockwise is positive.  (c) Critical values of the EM map consist of a conical lemon boundary surface and an enclosed thread.}
\label{theory}
\end{figure}

The actual calculation of the set of action-angle variables is highly nontrivial for all but the simplest integrable systems.  When monodromy is present, this construction is even more difficult as the action-angle variables are not defined globally.  Instead of constructing these variables directly, the question of whether or not a system contains monodromy is typically answered through a topological analysis of the system's so-called energy-momentum map (EM).
The EM takes points from $P^{2N}_{q,p}$ and maps them to the N-dimensional space made up of the constants of motion, denoted by $C^{N}_{F}$.  In general, $C^{N}_{F}$ contains two types of points, regular values and critical values.  Regular (critical) values correspond to motions where the differentials $dF_{k}$ are linearly independent (dependent).  Examples of motions corresponding to critical values are equilibria, with no motion at all, and relative equilibria, where motion occurs but the constants of motion are interdependent.  For the elastic pendulum, relative equilibria include pure springing and circular swinging at a fixed height.  As a general rule, the physically accessible domain of $C^{N}_{F}$ is bounded by critical values as the constants of motion are necessarily inter-dependent at extreme types of motion.  Monodromy can occur when critical values are imbedded inside the physically accessible domain of $C^{N}_{F}$.
For a concrete example of a region of $C_{F}^{N}$ both bounded by and containing imbedded critical values, see Fig. \ref{theory}(c).  Imbedded critical values allow for the existence of loops of regular values that cannot be continuously shrunken to a point.  After going once around one of these singularity-enclosing loops of initial conditions, the final action-angle variables will differ from the initial ones.  Determination of whether or not a system contains monodromy then reduces to finding the constants of motion and analyzing the locations and types of critical values for the EM.  We now summarize this calculation for the resonant elastic pendulum, the details of which are in Ref. \cite{Dullin04}.

\begin{figure}
\includegraphics[width=4cm,height=3.0cm]{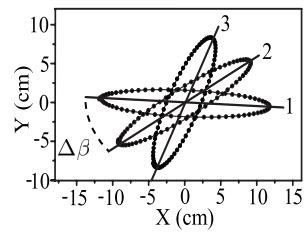}
\caption{Measured mass positions as projected onto the XY plane during three successive near-pure swinging motions.  These data are fit to the expected elliptical functional form.  The major axes of these ellipses represent the orientation of the swing plane $\beta$. In the example data, $\Delta \beta = 32^{\circ}$.}
\label{ellipses}
\end{figure}

The three-dimensional (3D) elastic pendulum consists of a bob of mass $m$ attached by a spring of constant $k$ to a pivot point and placed in a vertical gravitational field with acceleration $g$, as shown in Fig. \ref{system}.  Arbitrary initial conditions can result in either regular or chaotic motions \cite{Nunez94}.  We are interested in regular motions resulting from small displacements from the equilibrium-hanging position.  The two relevant modes of oscillation inherent to the system are swinging (pendular) motions with angular frequency $\omega_{p}=\sqrt{g/l}$, and springing motions with angular frequency $\omega_{s}=\sqrt{k/m}$, where $l$ is the equilibrium-hanging length.  These two frequencies are coupled because $l$ depends on $k$ via $l=l_{0}+\frac{mg}{k}$, where $l_{0}$ is the unstretched length of the spring.  Particularly interesting dynamics occur if the pendulum's swinging and springing frequencies are in a ratio 1:2, in which case energy can transfer efficiently between the two modes.  This transfer is an example of parametric resonance \cite{Vitt33}.  Specifically, nearly pure springing motion evolves into nearly pure swinging motion.  The projection of this near-pure swinging motion onto the XY plane is an ellipse with near-constant major-axis orientation.  The plane defined by this major axis and $\hat{z}$ is called the swing plane.  Energy then transfers back into the springing mode, and this cycle repeats.  The most striking aspect of this motion is that the  major axis of the projected ellipse precesses in a stepwise manner between successive near-pure swinging motions (Fig. \ref{ellipses}).  Furthermore, the step size between the swing planes is constant and a function of the initial conditions.  Theoretical treatments of this behavior have been carried out by Refs. \cite{Dullin04,Holm02,Lynch02}.  The main result in Ref. \cite{Dullin04} is that the stepwise precession $\Delta \beta$ of the swing plane orientation $\beta$ is a rotation number of the integrable approximation.  It is given by a smooth but multivalued function as a direct physical consequence of the existence of monodromy.  This function is given in terms of the constants of motion, which we now define.

The construction of the set of constants of motion begins with an examination of the Hamiltonian.  With the origin at the pivot point, the full Hamiltonian for the 1:1:2 resonant elastic pendulum is given by
\begin{equation}
\tilde{H}=\frac{1}{2}(p_{x}^{2}+p_{y}^{2}+p_{z}^{2})+z+\frac{1}{2}\nu^{2}(1-\frac{1}{\nu^{2}}-r)^{2},
\end{equation}
where the unit scaling $m=g=l=1$ has been used, $\nu=(\omega_{s}/ \omega_{p})=(k l / m g)^{1/2}=2$ is the ratio of springing to swinging frequencies, $r=\sqrt{x^{2}+y^{2}+z^{2}}$, and the $p_{i}$ are the $i^{th}$ dimensional momenta.
$\tilde{H}$ is invariant under a rotation around the z-axis.  Therefore the angular momentum $L_{z}=xp_{y}-yp_{x}$ is conserved.  After an expansion to cubic order about the free-hanging equilibrium position and an averaging over the fast dynamics, the effective Hamiltonian is given by
\begin{equation*}
\begin{split}
H & =\frac{1}{2}\left(p_{x}^{2}+p_{y}^{2}+p_{z}^{2} +x^{2}+y^{2}+\nu^{2}z^{2}\right) \\
& + \left(-\frac{\mu}{8}\left[\left(x p_{x}+y p_{y}\right)p_{z}+\left(x^{2}+y^{2}\right)z-\left(p_{x}^{2}+p_{y}^{2}\right)z\right]\right) \\
 & \equiv H_{2}+H_{3},
 \end{split}
\end{equation*}
where $\mu=\frac{\sqrt{2}}{16}(\nu^{2}-1)$, $H_{2}$ ($H_{3}$) is the quadratic (cubic) contribution, and $z$ is now measured from the equilibrium hanging position.  Under this approximation, the system is integrable with three constants of motion $\{H,H_{2},L_{z}\}$ in three degrees of freedom.
The critical values of the resulting $EM:\{x,y,z,p_{x},p_{y},p_{z}\}\mapsto \{H,H_{2},L_{z}\}$ form a ``conical lemon'' surface and a thread [Fig. \ref{theory}(c)].
The thread corresponds to purely springing motions at various total energies.  Loops in the 3D range of the EM are somewhat difficult to visualize, so a reduction to an equivalent two-dimensional (2D) system is carried out.  Each set of constants of motion is classified first by its value for $H_{2}$, effectively taking a slice out of the conical lemon in Fig. \ref{theory}(c).  A scaling of the remaining two quantities $\{H,L_{z}\}$ is implemented to map the various slices onto each other.  This mapping allows different sets of constants of motion to be compared easily as points in the now 2D lemon [Fig. \ref{theory}(a)].  The new dimensionless constants of motion are
\begin{equation}
\chi=\frac{H-H_{2}}{\mu H_{2}^{3/2}}, \quad  \quad \lambda=\frac{L_{z}}{H_{2}}.
\end{equation}
With this scaling, the thread pierces the 2D lemon at $(\lambda ,\chi)=(0,0)$, producing a singularity [Fig.\ref{theory}(a)]. This singularity is responsible for the existence of monodromy.

The presence of the monodromy-producing singularity causes a rotation number of the integrable approximation to be multivalued.  This rotation number corresponds to the step size $\Delta \beta$ of the stepwise-precessing swing plane during a full cycle of swinging, springing, and back to swinging.  Thus the physical consequences of monodromy are easily observed.  To first order, $\Delta \beta$ takes the form
\begin{equation}
\Delta \beta=arg(\chi+i \lambda).
\label{step}
\end{equation}
The arg function extracts the argument (phase angle) of a complex number and can be made single-valued but discontinuous through a branch-cut or can be viewed as multivalued and smooth.  The presence of this multivalued rotation number proves that monodromy exists in this system.  Explicit calculations for the various loops of constants of motion appear in Fig. \ref{theory}.  Curves in Fig. \ref{theory}(b) are labeled on the right with their corresponding loop numbers in (a).  Loops not enclosing the singularity return smoothly back to their initial values.  Loop 5 is different in that it does not.

The experimental goal is to measure $\Delta \beta$ at positions along loops similar to those in Fig. \ref{theory}(a) by varying initial conditions.  Successful experimental measurement of $\Delta \beta$ relies on relatively pure and long-lived swinging motion.  The former is necessary to determine when the motion is to be classified as purely swinging, purely springing, or in transition.  The latter allows for one or more complete ellipses to be traced out for each purely swinging motion, which facilitates a determination of the swing-plane orientation. The purity and lifetime of the swinging motion depends both on initial conditions and on how well the 2:1 resonance condition is satisfied. Our experimental parameters are $k=6.8(2)$ N/m, $m=0.224(1)$ kg, and unstretched spring length of $l_{0}=1.00(1)$ m.  These parameters yield $\nu=2.0(1)$, and pure swinging motions that persist for several seconds.  Typical energy-damping times are on the order of minutes.

\begin{figure}
\includegraphics[width=5.5cm]{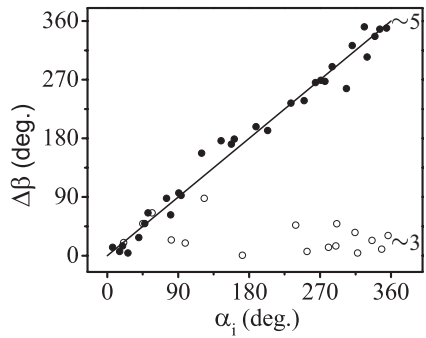}
\caption{Experimental measurements of the step angle between successive swinging motions as loops in ($\lambda,\chi$) space are mapped out.  A loop enclosing the singularity (solid circles) does not return to its initial value, thus demonstrating monodromy.  These data agree with the theoretical prediction (solid line).  A loop [loop 3 in Fig. 2(c)] not enclosing the singularity (open circles) is shown for comparison.}
\label{experimental}
\end{figure}

Once motion with well-defined swinging motions can be reproducibly created, the next requirement is an accurate determination of the ball's 3D position as a function of time, $\vec{x}$(t).  These data allow for a determination of the swing planes and therefore the stepwise precession angle, as well as the calculation of $\lambda$, $\chi$, and the expected $\Delta \beta$.  We capture the motion using two video cameras operating at 30 frames per second, one (XY camera) shooting video from below and the other (XZ camera) from the side.  To determine the position of the mass, a circle is fit to the ball's image for each frame
in the video.  Our circle-fitting scheme relies on the color gradient at the edge of the ball's image and therefore necessitates a high-contrast ratio between the ball and its background.  To achieve the needed contrast, the ball is painted
white and illuminated by intense halogen lights against a black background.
From measured $\vec{x}$(t), we fit the projected elliptical motion of the ball, as viewed by the XY camera during pure swinging motion, to an ellipse using a direct least-squares scheme.
The major axis defines the swing plane.  Fitting of successive pure-swinging motions yields the stepwise precession.  Our procedure is to measure three such swing planes and take the average of the two steps.  Typical data for a single experimental run are shown in Fig. \ref{ellipses}.  Each point represents the fitted position of the pendulum bob for a single frame of the captured video. Step angles between swing planes $\{1,2\}$ and $\{2,3\}$ typically differed by $2^{\circ}$ to $10^{\circ}$, depending on the ellipse eccentricity.
Instantaneous positions and velocities were measured during the part of a projected ellipse with the least curvature to minimize errors due to accelerations.  These instantaneous conditions were used to calculate the constants of motion and the expected step angle.

Step-precession behavior for two experimental loops is shown in Fig. \ref{experimental}.  As in Fig. \ref{theory}(b), the plots are denoted on the right by their approximate loop labels corresponding to Fig. \ref{theory}(a). Since swing planes are only determined within an additive constant of $180^\circ$, continuity of nearby $\alpha_{i}$ measurements is used to determine absolute positions of the solid circles.  Deviation of the singularity-enclosing loop ($\sim 5$, solid points) from the expected straight line is due mainly to nonideal measurements of the constants of motion.  In contrast, the noisy structure in the non-singularity-enclosing loop ($\sim 3$, open points) is a combination of these nonideal measurements and the difficulty of experimentally launching the pendulum bob with predetermined constants of motion.  This difficulty causes the experimental loops to not be perfect circles, but, crucially, they are still homotopic to circles.  This difficulty is not an issue for the singularity-enclosing loop, as it shares the same origin as the arg function.  The qualitative difference between the singularity-enclosing loop ($\sim 5$) and the non-singularity-enclosing loop ($\sim 3$) is clearly evident.  We see that the former loop does not come around to its initial value upon returning to the initial point, while the latter does.

It is a simple distinction: does or does not the size of a stepwise precession advance by $360^\circ$ as one maps out the behavior along a loop through constants-of-motion space? Yet, the key to our first-ever experimental study of classical monodromy is our ability to observe both sorts of behavior in the resonant elastic pendulum. We hope that this simple classical example can be  part of a solid foundation upon which to build the intuition necessary to understand the subtle, but by no means rare, instances in which monodromy profoundly influences the quantum spectra of atoms, molecules, and more complicated objects.

The authors would like to thank Sarah Anderson for the initial apparatus setup and Eric Cornell for many stimulating discussions.  Funding was provided by the Alfred P. Sloan Foundation and NSF.

\end{document}